# NEXT GENERATION EPICS INTERFACE TO ABSTRACT DATA[*]

J. Hill, LANL, Los Alamos, NM 87544USA
R. Lange, BESSY, 12489 Berlin, Germany

## Abstract

The set of externally visible properties associated with process variables in the Experimental Physics and Industrial Control System (EPICS)[1] is predefined in the EPICS base distribution and is therefore not extensible by plug-compatible applications. We believe that this approach, while practical for early versions of the system with a smaller user base, is now severely limiting expansion of the high-level application tool set for EPICS. To eliminate existing barriers, we propose a new C++ based interface to abstract containerized data. This paper describes the new interface, its application to message passing in distributed systems, its application to direct communication between tightly coupled programs co–resident in an address space, and its paramount position in an emerging role for EPICS — the integration of dissimilar systems.

## 1 IN PURSUIT OF HIGH LEVEL APPLICATIONS

The EPICS software was originally designed to be tool based approach to process control and this continues to be its primary application. However, the collaboration recognizes the benefits that might arise from more vigorous development of modular advanced physics modeling and control toolkits that are closely integrated with EPICS. There may be cultural and geographical obstacles to this type of open source collaboration, but our experience makes us suspect that these barriers are not insurmountable, and if not then perhaps our aspirations for more efficient development of advanced toolkits will be fulfilled if certain technical limitations in the existing EPICS communication software interfaces are eliminated.

## 2 FUNDAMENTALS WE DON'T INTEND TO CHANGE

Certain aspects of the existing EPICS communication software interfaces appear to be important facilitators for advanced toolkits. Close integration with process control systems requires efficient publish-and-subscribe communication strategies. Message-batching capabilities also improve communication efficiency. Software interfacing with systems capable of independent actions needs interfaces that can generate an asynchronous response synchronized with external events. An infrastructure that encourages proper design of distributed software systems is also important. For example, in multi-threaded distributed systems, toolkits need communication software interfaces designed to avoid application programmer introduced mutual exclusion deadlocks. Interfaces must also be properly structured to encourage robust response to loss of communication or other hardware resources. Portability between workstations and embedded systems is an important requirement for certain advanced applications. These capabilities are required by process control components. We expect that they will also be beneficial to advanced modeling and control toolkits.

## 3 FOSTERING INTEGRATION WITH HIGH LEVEL APPLICATIONS

Several physics modeling and control toolkits have been successfully interfaced with EPICS. These programs are not shared between sites as frequently as we had originally hoped, and their view of EPICS tends to be a fairly narrow one where EPICS is only a source and destination for data. In our experience the fundamental requirement for vigorous open-source software collaboration is well-defined software interfaces that break a large software effort into a system of moderate sized modular replaceable components. Unfortunately, while the EPICS software interfaces satisfy the fundamental communication requirements for distributed systems, they are lacking capabilities encouraging collaborative layering of software modules above and beyond the requirements of distributed process control.

The fundamental endpoint for communication within EPICS is an abstract "process variable" with the built-in set of properties listed in Table 1.

Table 1: Process Variable Properties

| Name | Display limits |
|---|---|
| Class | Control limits |
| Data type | Alarm limits |
| Vector dimension | Alarm condition |

[*] Work supported by the Office of Energy Research, Basic Energy Science of the US Department of Energy, the Oak Ridge National Laboratory, the Bundesministerium für Bildung, Wissenschaft, Forschung und Technologie (BMBF), and the Land Berlin.

| Value | Alarm acknowledge transient |
|---|---|
| Time stamp | Alarm acknowledge severity |
| Units | Number of decimal digits |
| Multi-state label names | |

Furthermore, EPICS clients can subscribe for process variable property updates to be sent when triggered by any combination of events from the built-in set listed in Table 2.

Table 2: Process Variable Subscription Events

| Change of state (default dead band) |
|---|
| Change of state (archiving dead band) |
| Alarm condition change of state |

Unfortunately, these built-in property and event sets are inadequate for integration of components that fall outside the realm of traditional process control.

For example, a data acquisition system might have an archiving engine that spools physics events off to disk. When a particular event occurs, a set of process variable properties must be gathered together and sent off to the archiving engine. In this context it is important to guarantee that we synchronize acquisition of all these property values with the specified triggering event. Currently, event and property sets are not extensible by components that plug-and-play with EPICS. Therefore, it is difficult to guarantee that a subscription update associated with one process variable is synchronized in any way with an independent subscription update associated with another process variable. Advanced toolkits need the capability to define new complex data types, and new event types, unknown to the system internals when they were compiled.

Considering another example, suppose that we have a high-level tool kit that wishes to be portable over a range of different astronomical telescopes. Suppose that this toolkit has two components: the star tracking system and the telescope positioning system. When the star tracking system needs to tell the telescope positioning system about a new position it must communicate at least two parameters. In the current EPICS system we can write to only one process variable at a time and therefore ad-hoc methods must be conceived which allow both parameters to be communicated before the telescope positioning system initiates the task of gently slewing the telescope to a new position. Otherwise, the telescope positioning system might initiate a move after receiving only one of the parameters risking a less than optimal path to its destination. Of course we can write the two position related process variables and then write to a third process variable that initiates the action. However, this approach does not foster the development of well-defined interfaces between modular high-level software components. Instead, we are left with a poorly enforced and error prone interface. The lack of multi-thread safety in this type of ad-hoc interface is of particular concern to a distributed control system. In contrast, when toolkits can install new complex data types initially unknown to core system components EPICS can accommodate modern software communication paradigms such as message passing and command completion synchronization.

## 4 INTERFACING WITH PROPRIETARY DATA — CURRENT PRACTICE

Many self-describing data file formats and their associated programming interfaces are available. However, in our experience there are two methods commonly in use by communications software systems for interfacing with arbitrary, complex structured, and application specific data.

With remote procedure call systems such as CORBA[2] there is a compiler that reads a source file with a specialized syntax describing data structures and any associated function call interfaces. This compiler generates a header file for the target language describing these data structures and interfaces. Object code stubs that can be used to transfer data on and off the wire are also produced. This approach is very efficient at runtime. However, it is not possible to extract an arbitrary subset of the elements within a compound data type, and therefore the communication system cannot arbitrarily map between data structures in different programs. This is a direct result of the communication system's being oblivious to the purposes of the fields in the user defined data structures. In publish and subscribe systems such as EPICS this limitation might impact flexibility and compatibility between modular components of the toolkit. This approach typically also has difficulties interfacing with array data when multi-dimensional bounds may change at run time.

In contrast, systems such as GDD[3] and CDEV[4] use a C++ class to encapsulate proprietary data. This approach stores the data internally as a union, or a linked list of unions if the data is compound. Each entry in the data is assigned a property name such as units, limits, or time-stamp. This allows extraction of an arbitrary subset of elements within a compound data type and installation of new elements into complex compound data at runtime. However, this introduces a large storage and execution overhead because knowledge of the data type's structure must be stored with every instance of the data. GDD provides mechanisms to efficiently index data using its property identifier, but considerable confusion has resulted from these capabilities being available only in certain modes of operation. This approach also requires a fairly large amount of code in its implementation. Users appear to find interfacing with this approach daunting[5], probably because they must constantly convert between their native storage formats and the communication system's imposed data container.

## 5 INTERFACING WITH PROPRIETARY DATA — ANOTHER APPROACH

We identify a third distinct approach to interfacing communications systems with proprietary data. With this approach there is a C++ abstract base class (an interface) that is used to introspect the structure of the arbitrarily complex proprietary data. If a toolkit element chooses to export it's proprietary data using this interface, then any programs that know the interface may examine or modify the data. A small support library provides functions for comparing, converting, and copying between dissimilar data sets. The toolkit element is not required to store its data in any particular format or organization. Nevertheless, knowledge of a complex data type's structure can be determined at compile time, and therefore access to the data can be efficient.

All data exported through this interface is assigned a property name. A property name may be "weight", "units", "maximum", or potentially any name that a group of programs mutually agree upon. A set of data with unique property names may be stored in a container that must also be assigned a property name. Properly interfaced data must provide a traversal function exporting knowledge of the purpose, the primitive data type, and the vector bounds of each participating property. When a toolkit element needs to extract a property subset out of an arbitrary data container it requires capabilities that efficiently locate specific properties in an unknown container. Therefore, properly interfaced data must also provide a function that locates a particular property, and library functions are provided to assist with efficient implementation. Compared to the traversal mechanism, we expect to introduce the additional flexibility required by certain applications at the expense of some loss of runtime efficiency.

Compared to CDEV and GDD this approach is less complex, because the data is not transformed into a new storage format when it crosses the interfaces of the communication system. This reduces the size of the support libraries and the toolkit elements. Storage overhead can also be lower than with GDD and CDEV because the description of the data may, at the users discretion, be stored separately from each data instance.

Compared to remote procedure call systems, we do not need to write a compiler that generates object code stubs for moving data on and off the wire. The stubs are more efficient, but we expect that the additional overhead will not be significant in this context. The proposed approach can introduce similar per instance storage overhead compared to remote procedure call systems, but these systems do not include facilities to extract a subset of properties from a properly interfaced arbitrary data structure. Finally, we observe that this approach can be used to efficiently interface to either of the above two approaches, but the opposite is not possible for a traditional remote procedure call system such as CORBA.

## 6 CONCLUSIONS

EPICS includes a comprehensive set of communication primitives that are essential for distributed process control, but we aspire to cultivate advanced integration of high-level modular toolkits. The fundamental endpoint for communication within EPICS is an abstract "process variable" with a fixed set of named properties and subscription update events. Advanced toolkits need the capability to define new complex data types and new subscription update events unknown to the system internals when they were compiled. To eliminate existing barriers, we propose a new C++ based interface to abstract containerized data. The new interface was compared to existing practice revealing important distinctions. A subset of properties can be extracted from a properly interfaced proprietary data set. The interface does not impose a storage format, but nevertheless knowledge of an arbitrary data type's structure can be efficiently determined at compile time.

---

[1] http://www.aps.anl.gov/epics/